\documentclass[a4paper,11pt]{article}
\usepackage{pos}
\usepackage[capitalize]{cleveref}
\usepackage{wrapfig}

\title{Predictions of Neutrino Fluxes at the Forward Physics Facility at the High-Luminosity LHC}
\ShortTitle{Neutrino Fluxes at the Forward Physics Facility}

\author*[a]{Dennis Soldin}
\author[a]{Isabella Coronado}
\author[a]{Elijah Waters}
\author[b,c]{Felix Kling}

\affiliation[a]{Department of Physics and Astronomy, University of Utah, Salt Lake City, UT 84112, USA}
\affiliation[b]{Deutsches Elektronen-Synchrotron DESY, 22607 Hamburg, Germany}
\affiliation[c]{Department of Physics and Astronomy, University of California, Irvine, CA 92697, USA}

\emailAdd{dennis.soldin@utah.edu}

\abstract{
High-energy collisions at the Large Hadron Collider (LHC) have traditionally focused on particle production at small pseudorapidities. However, to further utilize the valuable data from particles produced at the ATLAS interaction point along the beamline, the proposed Forward Physics Facility (FPF) aims to study particle production in the far-forward region at the high-luminosity LHC. The FPF will house a suite of experiments with the ability of enhancing hadronic interaction models by measurement of the resulting neutrino fluxes. These advancements are critical for astroparticle physics, linking forward scattering processes to extensive air showers and improving our understanding of cosmic-ray interactions. This work investigates simulated neutrino fluxes in the far-forward region from proton-proton collisions at ATLAS, analyzing final state particles propagated to this new facility. Results from this simulation provide theoretical expectations for the neutrino fluxes at the FPF, offering insights to refine hadronic interaction models, including the most recently developed, and better estimate atmospheric neutrino backgrounds in astrophysical neutrino telescopes.
}

\ConferenceLogo{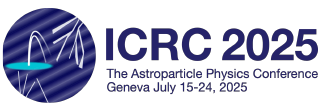}

\FullConference{39th International Cosmic Ray Conference (ICRC2025)\\
 15–24 July 2025\\
Geneva, Switzerland\\}

\begin{document}
\maketitle

\section{Introduction}

Interactions of high-energy cosmic rays with air nuclei in the atmosphere generate extensive air showers (EASs). Cosmic rays can only be studied indirectly through the detection of secondary particles from EASs by large ground-based detector arrays. To infer the properties of primary cosmic rays (such as their energy and momentum) from these measurements, simulations are required to interpret the EAS development. Modeling hadronic interactions in the forward region across a wide energy range remains a key challenge in advancing our understanding of EASs, since conventional collider experiments cannot directly probe this domain.

\begin{figure}[!b]
  \centering
  \vspace{-1.em}
  \includegraphics[width=0.9\textwidth]{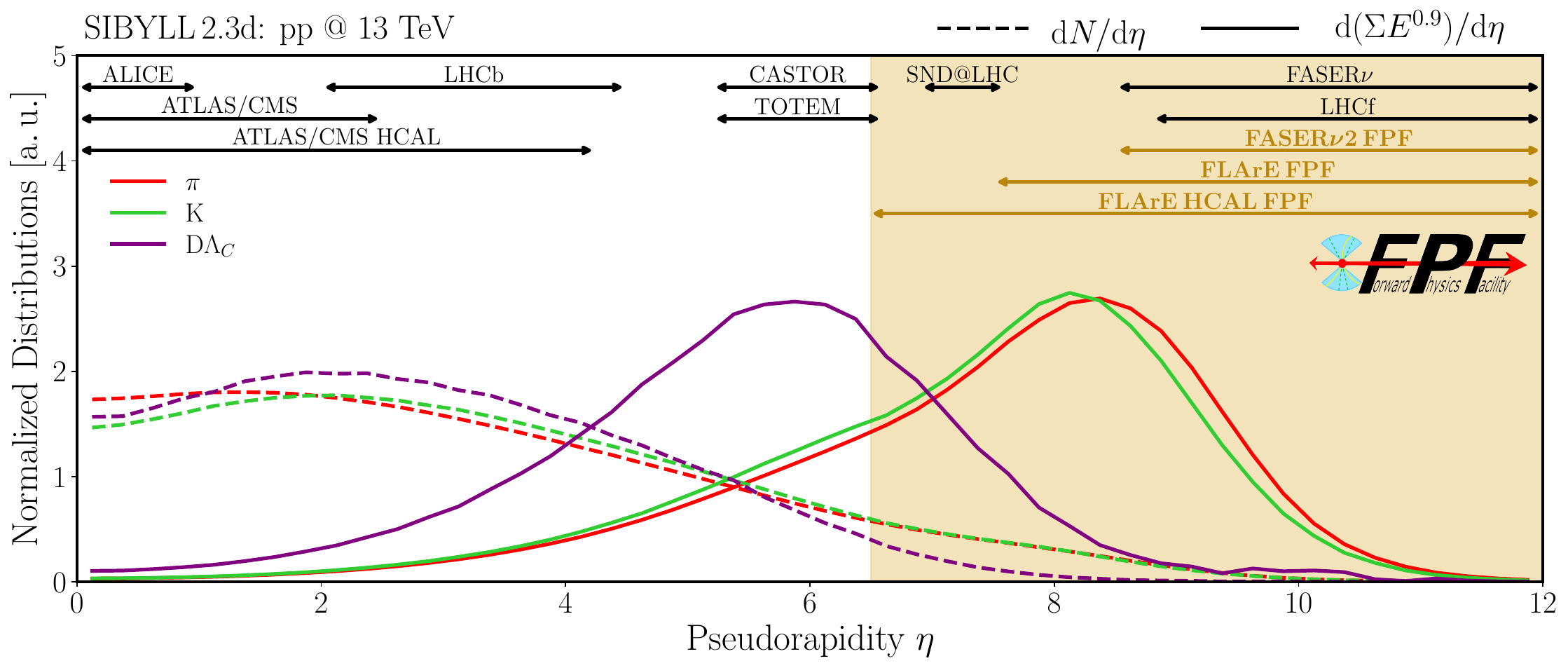}
  \vspace{-.5em}
  
  \caption{Simulated densities of particles as a function of their pseudorapidity~\cite{Albrecht:2021cxw} in arbitrary units (dashed lines) in proton-proton collisions obtained from EPOS-LHC~\cite{Pierog:2013ria}. Solid lines show the estimated number of muons produced by these particles, assuming an equivalent energy for the fixed target collisions in the laboratory frame, $E_{\rm{lab}}$, and $N_\mu \propto E_{\rm{lab}}^{0.9}$.}
  \label{fig:eta}
  \vspace{-1.em}
\end{figure}

\Cref{fig:eta} illustrates simulated particle densities from proton-proton collisions (solid lines), with pseudorapidity ($\eta$) ranges which are covered by existing LHC experiments overlaid~\cite{Albrecht:2021cxw}. Except for LHCb, the LHC experiments were not designed to perform precision tracking and particle identification at forward rapidities. The dashed lines represent the estimated number of muons, $N_\mu$, produced by different hadrons, assuming an equivalent energy, $E_{\text{lab}}$, for the fixed target collisions in the laboratory frame and $N_\mu \propto E^{0.9}$. The mid-rapidity region only has a minimal effect on the production of particles in EASs, which are predominantly emitted in the forward region (\(\eta > 4\)). Their production is mainly driven by hadron-ion collisions at low momentum transfer, which cannot be described in the context of perturbative QCD. Given the difficulty in modeling hadron production from first principles, combined with the lack of data from current collider experiments, EAS simulations typically rely on phenomenological interaction models. However, discrepancies between model predictions and experimental data have been reported by several cosmic-ray experiments over many years, a phenomenon referred to as the muon puzzle in EASs. In particular, measurements at the Pierre Auger Observatory show an excess in the number of muons compared to simulations~\cite{PierreAuger:2014ucz, PierreAuger:2016nfk,PierreAuger:2024neu} and a meta-analysis of data from nine air-shower experiments has revealed an energy-dependent trend in these discrepancies~\cite{Dembinski:2019uta,Soldin:2021wyv,Cazon:2020zhx}. Recent studies suggest that these discrepancies indicate a severe deficits in our understanding of particle physics~\cite{Albrecht:2021cxw,ArteagaVelazquez:2023fda}. Thus, accurate measurements of hadronic interactions, particularly in the far-forward region, are crucial for validating and refining current EAS models.

\section{The Forward Physics Facility} 
\label{sec:FPF}

The proposed Forward Physics Facility (FPF) \cite{Anchordoqui:2021ghd,Feng:2022inv,Adhikary:2024nlv} at the high-luminosity LHC (HL-LHC) is designed to take advantage of the intense far-forward flux produced at colliders. In 2024, the FASER and SND@LHC collaborations reported the first measurement of high-energy collider neutrinos~\cite{FASER:2023zcr,SNDLHC:2023pun} and demonstrated the capability of forward neutrino experiments to study hadronic interaction models with high precision~\cite{FASER:2024ref}. The FPF is expected to further extend the forward physics program at CERN into the HL-LHC era to investigate particle production in the high pseudorapidity region (see \cref{fig:eta}) with unprecedented statistics, and to perform neutrino physics studies that bridge collider and astroparticle physics. 

\begin{figure}[!b]
    \centering
    \includegraphics[width=1\linewidth]{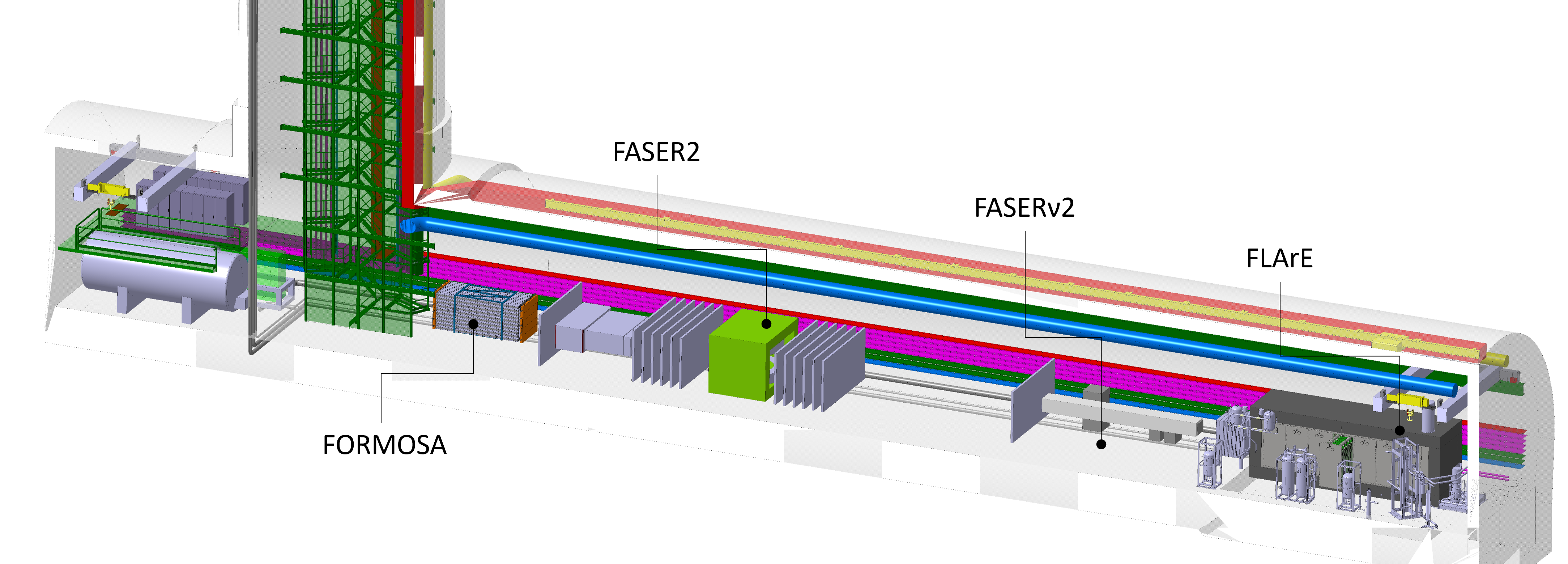}
    \caption{Layout of the FPF~\cite{Adhikary:2024nlv}, located around 620\,m from the ATLAS interaction point and housing the four proposed experiments FASER2, FASER$\nu$2, FLArE, and FORMOSA. See text for details.}
    \label{fig:baseline}
\end{figure}

The FPF will be located 88\,m underground along the line-of-sight (LOS) of the ATLAS interaction point, at a distance of approximately 627\,m, and shielded by over 200\,m of rock. The facility, shown in \cref{fig:baseline}, will span roughly 75\,m in length and 11.8\,m in internal width, providing the necessary infrastructure to host a diverse set of experiments aimed at investigating physics phenomena at pseudorapidities above $\eta\simeq7.5$. The energies of neutrinos reaching the FPF will range from a few 10\,GeV up to several TeV during HL-LHC operations. The experiments are designed to reconstruct charged particle tracks from neutrino interactions and determine all three  neutrino flavors at energies comparable to those in EAS, but under controlled laboratory conditions.

\subsection{Experiments} 

Currently, four experiments are proposed to be housed in the FPF, as shown in \cref{fig:baseline}. Each experiment is based on different detector technologies and optimized for a specific purpose, as briefly summarized in the following. A comprehensive review can be found in Refs.~\cite{Feng:2022inv,Adhikary:2024nlv}.

\vspace{-0.6em}

\begin{itemize}
\itemsep0.07em 

\item \textbf{FASER2:} An upgraded version of the FASER experiment~\cite{FASER:2018ceo}, a magnetic tracking spectrometer designed for the detection of neutrinos, as well as long-lived particles, such as dark photons, axion-like particles, heavy neutral leptons, among other beyond Standard Model scenarios, and to distinguish $\nu$ and $\bar\nu$ charged current scattering in the FPF neutrino detectors by measuring the muon charge. It includes an enlarged volume and enhanced tracking and calorimetry systems relative to its predecessor. 

\item \textbf{FASER$\nu$2:} Successor to the experiment FASER$\nu$ \cite{FASER:2020gpr} and a dedicated neutrino detector capable of observations of all three neutrino flavors. It uses dense emulsion layers interleaved with tungsten plates to enable precise reconstruction of neutrino interactions. With its excellent spatial resolution, it will not only observe neutrino interactions but will also distinguish between neutrino flavors and measure their energy distributions.

\item \textbf{FLArE:} A versatile and large-scale (10 ton) liquid argon time projection chamber, which reconstructs three-dimensional particle tracks by measuring the ionization from charged particles as they pass through the detector. This experiment, also covering the lower rapidity region, will detect neutrinos of all flavors and provide an accurate reconstruction of the interaction final states. 

\item \textbf{FORMOSA:} Unlike the neutrino detectors, this experiment is highly focused on beyond Standard Model searches. It has world-leading sensitivity to detect millicharged particles via arrays of scintillator panels and coincidence timing techniques.
\end{itemize}

\vspace{-0.5em}

This suite of experiments will comprise a comprehensive forward physics program~\cite{Feng:2022inv}, providing measurements of collider neutrinos with unprecedented statistics (up to millions of neutrinos) and high precision during the HL-LHC era. The normalization of the muon-neutrino flux, for example, can be measured at the per-mille level at the FPF. Under this program, crucial data will be collected with the ability to further build a connection between collider and astroparticle physics.

\section{Neutrino Fluxes at the FPF} 
\label{sec:fluxes}

To predict the neutrino fluxes at the FPF, a multi-step simulation is used that models proton-proton collisions at the LHC, propagates particles from the ATLAS interaction point toward the FPF, and simulates hadron decays into neutrinos, as well as their interactions in the FPF detectors. This simulation procedure and the resulting neutrino flux predictions are discussed in the following.

\subsection{Event Generators} 

The software tool CHROMO~\cite{dembinski_chromo_2023} is used to generate proton-proton collisions at the \mbox{ATLAS} interaction point with various hadronic interaction models. 100,000 collisions at $\sqrt{s} = 14$\,TeV are generated for each of the hadronic interaction models SIBYLL\,2.3d~\cite{Riehn:2019jet}, DPMJET-III~\cite{Roesler:2000he}, EPOS-LHC~\cite{Pierog:2013ria}, EPOS-LHC-R~\cite{Pierog:2023ahq}, QGSJET-II-04~\cite{Ostapchenko:2013pia,Ostapchenko:2005nj}, QGSJet-III~\cite{Ostapchenko:2024myl}, and Pythia 8.2~\cite{Sjostrand:2014zea}, including a dedicated forward tune~\cite{Fieg:2023kld}. The model SIBYLL$^\bigstar$~\cite{Riehn:2024prp} is a series of phenomenologically modified versions of Sibyll 2.3d in order to increase muon production from hadronic multi-particle production processes which provide a possible solution to the muon puzzle. This is done by enhancing the production of baryon-antibaryon pairs, kaons, or neutral rho mesons. It also provides a mixed model with baryon-antibaryon and neutral rho meson enhancements. These enhancements take effect at energies above $\sqrt{s}=13$\,TeV to remain within bounds from accelerator measurements, however, in order to study potential effects that can be observed at the FPF, we lower this threshold to $\sqrt{s}=10$\,TeV. Another example accounting for enhanced kaon production can be realized in a simple toy model by allowing the substitution of pions with kaons in SIBYLL-2.3d at large pseudorapidities ($\eta > 4$)~\cite{Anchordoqui:2022fpn}. However, this model typically results in even larger kaon production than the corresponding SIBYLL$^\bigstar$ version and is not further considered in this work.

\subsection{Fast Neutrino Flux Simulation}

The propagation of all particles, their decay into neutrinos, as well as the interaction of neutrinos in the FPF detectors is simulated using dedicated software~\cite{Kling:2021gos}. It has been shown that neutrinos, especially at the higher energies of interest, are primarily produced in the vacuum of the LHC beam pipe, while neutrinos from secondary interactions only contribute a very small fraction to the neutrino flux. The fast simulation obtains the trajectories of hadrons through the LHC beam pipe and magnets, using a geometrical model used from BDSIM~\cite{Nevay:2018zhp}, and models their decay into neutrinos, following Pythia~8.2~\cite{Sjostrand:2014zea}. The interactions of neutrinos in the FPF detectors is modeled based on GENIE~\cite{Andreopoulos:2015wxa}. The entire simulation procedure is  implemented as a RIVET~\cite{Buckley:2010ar} module which provides easy access to the entire community. For the purpose of illustration, we simulate neutrino fluxes in FASER$\nu$2 using this approach, however, the general conclusions are mostly independent of the specific FPF experiment.

\begin{figure}[!b]
    \centering
    \vspace{-0.5em}
    \includegraphics[width=1\linewidth]{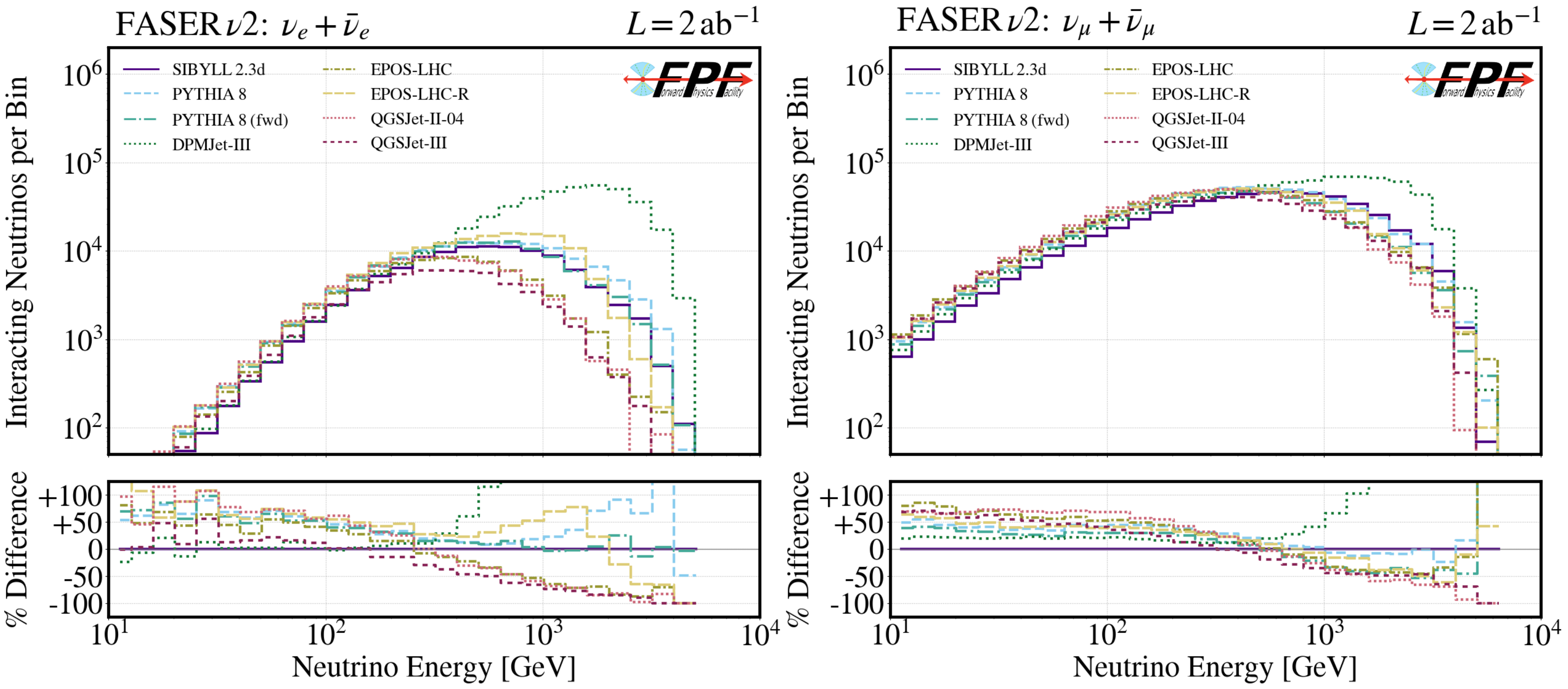}
    \caption{Simulated neutrino energy for electron neutrinos (left) and muon neutrinos (right) interacting in FASER$\nu$2, obtained from different hadronic interaction models. The bottom panel shows the corresponding model differences with respect to SIBYLL\,2.3d.}
    \label{fig:totals}
    \vspace{-0.5em}
\end{figure}

\subsection{Results and Discussion}  

\begin{figure}[tb]
    \centering
    \vspace{-1.em}
    \includegraphics[width=1\linewidth]{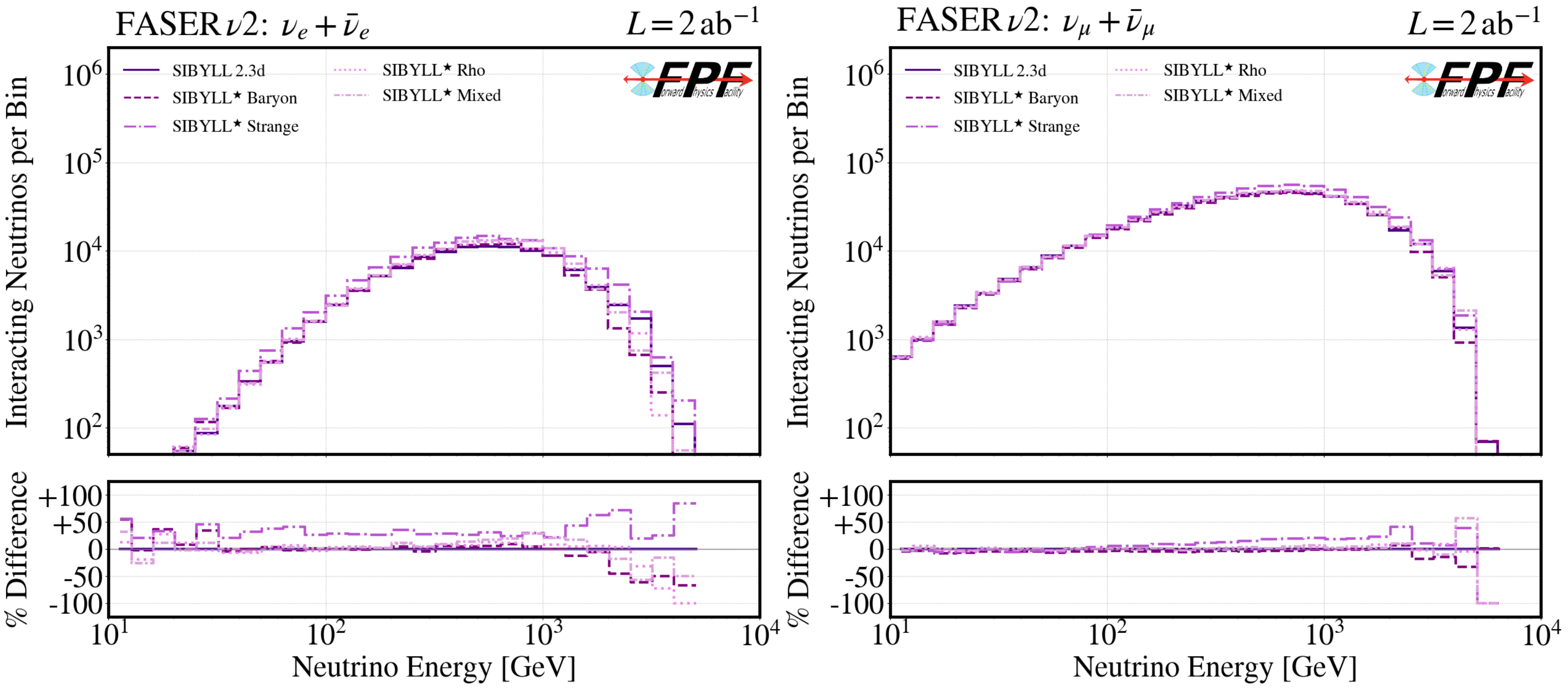}
    \caption{Simulated neutrino energy for electron neutrinos (left) and muon neutrinos (right) interacting in FASER$\nu$2, obtained from different variations of the SIBYLL$^\bigstar$ model. The bottom panel shows the corresponding model differences with respect to the benchmark model SIBYLL\,2.3d.}
    \label{fig:totals_SibStar}
\end{figure}

\Cref{fig:totals} shows the simulated neutrino energy spectra for electron neutrino and muon neutrino interactions in FASER$\nu$2, assuming an integrated luminosity of $2\,\mathrm{ab}^{-1}$. These predictions, are based on the interaction models SIBYLL\,2.3d, DPMJET-III, EPOS-LHC, EPOS-LHC-R, QGSJET-II-04, QGSJet-III, and Pythia\,8.2, including a forward tune. Also shown are the differences between the models with respect to SIBYLL\,2.3d which are up to 100\% at low energies, depending on the particular model, much larger than the expected uncertainties at the FPF~\cite{Kling:2021gos,Cruz-Martinez:2023sdv}. At the highest energies (above a few 100\,GeV), the model differences become even larger. This is mainly due to very different treatment of charm hadrons between the models. It is interesting to note that the differences are minimal at energies of around 300\,GeV. Differences in the predicted neutrino spectra can also be observed for predictions based on SIBYLL$^\bigstar$ at a similar level, in particular, for the kaon enhancement model, as shown in \cref{fig:totals_SibStar}. 

\begin{figure}[!b]
    \centering
    \includegraphics[width=1\linewidth]{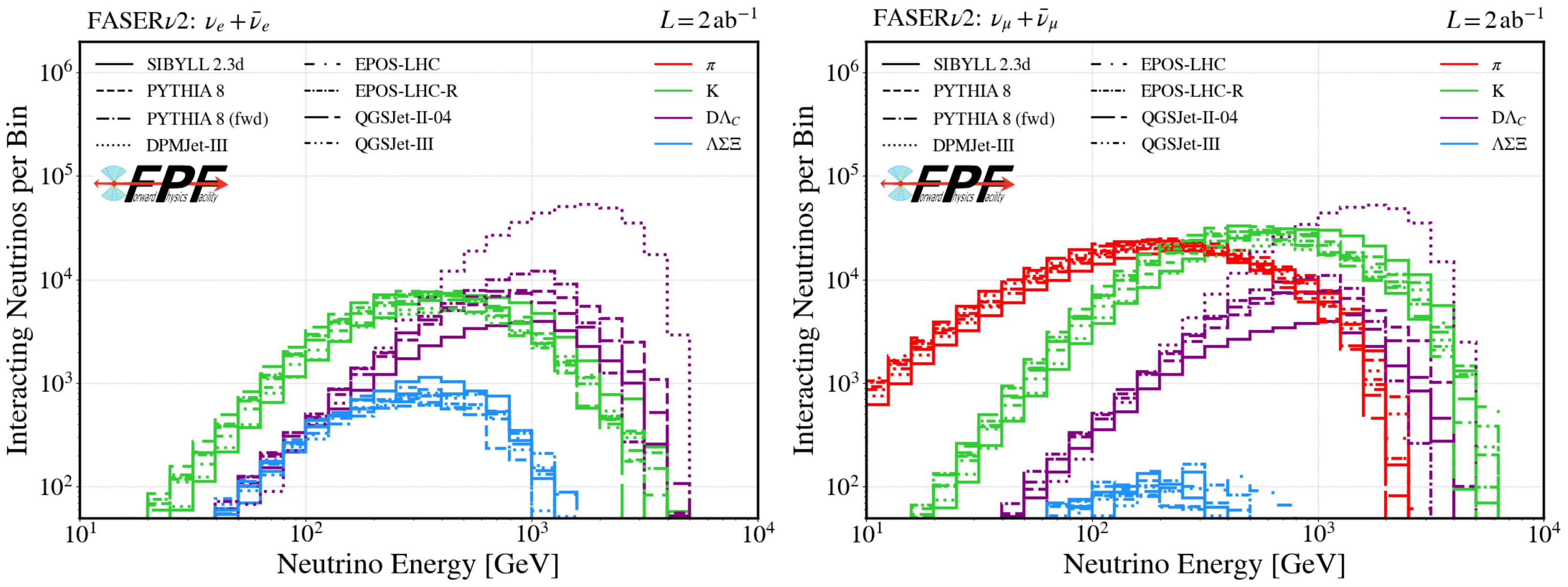}
    \caption{Simulated neutrino energy spectra for electron neutrinos (left) and muon neutrinos (right) interacting in FASER$\nu$2, obtained from various hadronic interaction models for different production modes, i.e., pion (red), kaon (green), charm (purple), and hyperon (blue) decays.}
    \label{fig:parent}
    \vspace{-1.em}
\end{figure}

\Cref{fig:parent} shows the corresponding composition of parent hadrons of neutrinos observed in FASER$\nu$2. The ratio of electron neutrino to muon neutrino fluxes offers an indirect method for determining the ratio of charged kaons to pions. As shown in the figure, neutrinos from pion and kaon decays have distinct energy spectra, which allows them to be differentiated, as recently demonstrated by FASER~\cite{FASER:2024ref}. Additionally, neutrinos from pion decays are more concentrated around the LOS compared to those from kaon decays. This is due to the lower mass of the pion and the resulting smaller transverse momentum. Therefore, the rapidity distribution of neutrinos can be additionally used to disentangle the origins of the neutrinos and provide an estimate of the pion-to-kaon ratio. In fact, it is expected that flux measurements at the FPF will be able to constrain the individual flux components with percent-level precision~\cite{Kling:2023tgr} and thereby provide important tests of hadronic interaction models.

\begin{wrapfigure}{r}{0.46\textwidth}
    \centering
    \vspace{-.5em}
    \includegraphics[width=1.\linewidth]{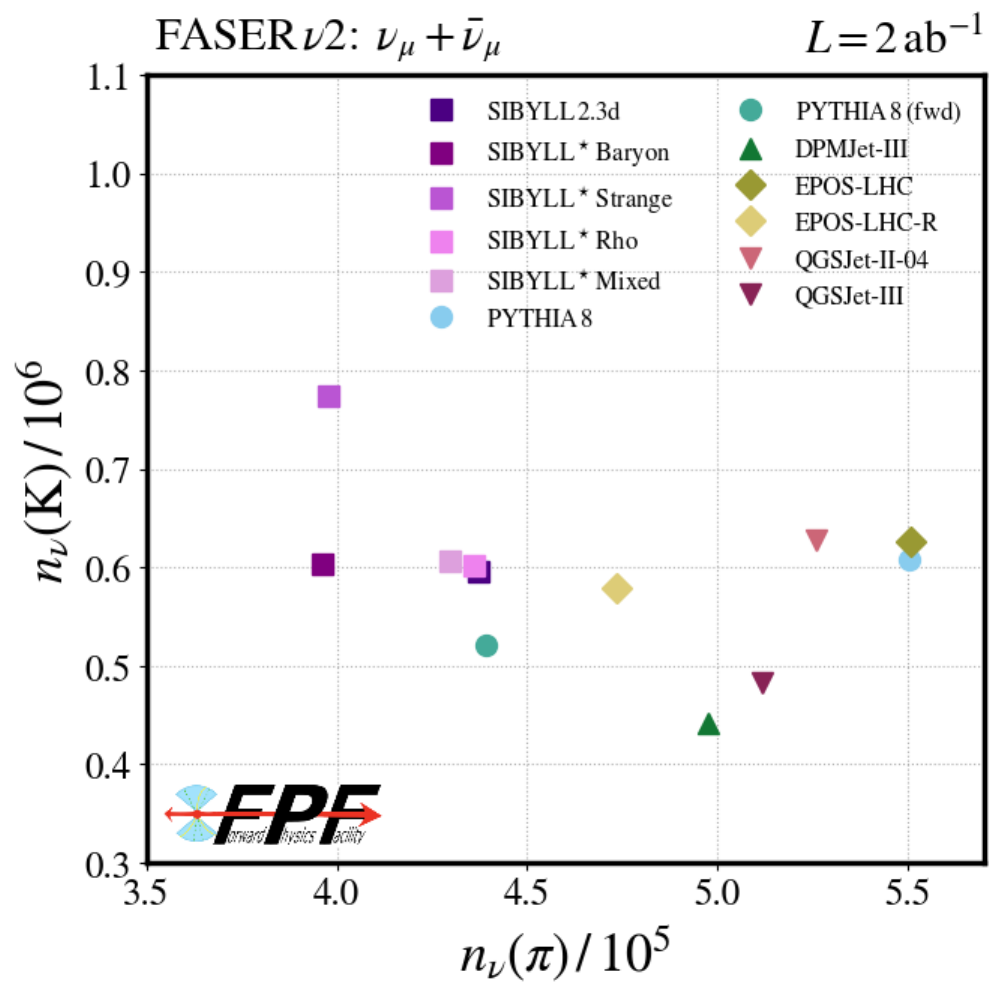}
    \caption{Number of muon neutrinos from kaon decays and pion decays, interacting in FASER$\nu$2, obtained from different hadronic models.}
    \label{fig:piK_scatter}
    \vspace{-2em}
\end{wrapfigure}

\begin{figure}[!b]
    \centering
    \vspace{-1em}
    \includegraphics[width=0.98\linewidth]{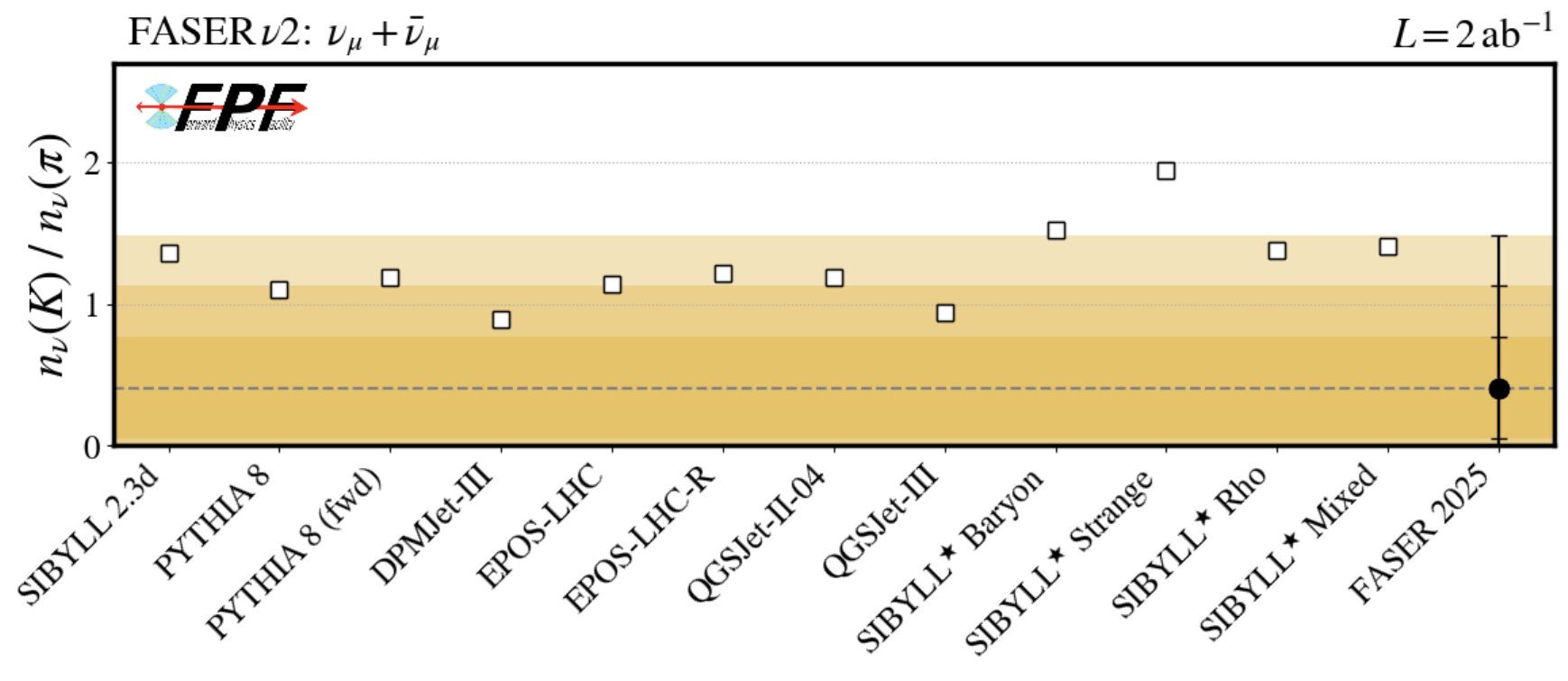}
    \vspace{-1em}
    \caption{Ratio of the number of muon neutrinos from kaon and pion decays in FASER$\nu$2, obtained from different models. Also shown is the measurement by FASER~\cite{FASER:2024ref} with an integrated luminosity of $(65.6\pm1.4)\,\mathrm{fb}^{-1}$, including 1$\sigma$, 2$\sigma$, and 3$\sigma$ uncertainty bands.}
    \label{fig:piK_ratio}
    \vspace{-1.em}
\end{figure}

The number of neutrinos from pion and kaon decays expected in FASER$\nu$2, obtained from the different hadronic interaction models, is shown in \cref{fig:piK_scatter}. Significant differences between various model predictions can be observed. This is further presented in \cref{fig:piK_ratio} which shows the ratio of the number of neutrinos from kaons to pions in comparison to a recent measurement by FASER~\cite{FASER:2024ref}. All hadronic model predictions are in tension with the experimental data, which indicates an overestimation of the ratio of kaons to pions in all models. However, it is important to note that FASER probes a narrower pseudorapidity region ($\eta>8.8$) compared to FASER$\nu$2 ($\eta>8.4$). A measurement of the individual flux components at the FPF with high statistics is therefore of particular importance to test current model predictions. 

\section{Conclusions}
\label{sec:conclusions}

We have presented simulations of the expected neutrino fluxes at the FPF, in particular, in the FASER$\nu$2 experiment, assuming various hadronic interaction models. These model predictions differ by up to 80-100\%, much larger than the anticipated uncertainties of the FPF experiments, and even more at the highest energies. The examples presented in this work demonstrate how the FPF will be uniquely positioned to test and constrain hadronic models, providing a significant advancement in our understanding of multi-particle production in EASs. Thereby, measurements at the FPF will lead to a better understanding of the properties of the highest-energy cosmic rays and will be critical for reducing associated uncertainties in high-energy astrophysical neutrino searches.

\section*{Acknowledgements}

We would like to thank Anatoli Fedynitch and Felix Riehn for useful discussions and invaluable technical support. We acknowledge the support and resources from the Center for High Performance Computing at the University of Utah. I.\,C. is supported by the Summer Undergraduate Research Fellowship Program at the University of Utah. 
F.\,K. is supported by the Heising-Simons Foundation Grant 2020-1840 and NSF Grants PHY-2210283 and PHY-2514888.


\bibliographystyle{ICRC}
\bibliography{references}

\providecommand{\href}[2]{#2}\begingroup\raggedright\begin{thebibliography}{10}

\bibitem{Albrecht:2021cxw}
J.~Albrecht {\em et~al.}, \href{http://dx.doi.org/10.1007/s10509-022-04054-5}{{\em Astrophys. Space Sci.} {\bfseries 367} no.~3, (2022) 27}.

\bibitem{Pierog:2013ria}
T.~Pierog {\em et~al.}, \href{http://dx.doi.org/10.1103/PhysRevC.92.034906}{{\em Phys. Rev. C} {\bfseries 92} no.~3, (2015) 034906}.

\bibitem{PierreAuger:2014ucz}
{\bfseries Pierre Auger} Collaboration, A.~Aab {\em et~al.}, \href{http://dx.doi.org/10.1103/PhysRevD.91.032003}{{\em Phys. Rev. D} {\bfseries 91} no.~3, (2015) 032003}.

\bibitem{PierreAuger:2016nfk}
{\bfseries Pierre Auger} Collaboration, A.~Aab {\em et~al.}, \href{http://dx.doi.org/10.1103/PhysRevLett.117.192001}{{\em Phys. Rev. Lett.} {\bfseries 117} no.~19, (2016) 192001}.

\bibitem{PierreAuger:2024neu}
{\bfseries Pierre Auger} Collaboration, A.~Abdul~Halim {\em et~al.}, \href{http://dx.doi.org/10.1103/PhysRevD.109.102001}{{\em Phys. Rev. D} {\bfseries 109} no.~10, (2024) 102001}.

\bibitem{Dembinski:2019uta}
{\bfseries EAS-MSU, IceCube, KASCADE Grande, NEVOD-DECOR, Pierre Auger, SUGAR, Telescope Array, Yakutsk EAS Array} Collaboration, H.~P. Dembinski {\em et~al.}, \href{http://dx.doi.org/10.1051/epjconf/201921002004}{{\em EPJ Web Conf.} {\bfseries 210} (2019) 02004}.

\bibitem{Soldin:2021wyv}
{\bfseries EAS-MSU, IceCube, KASCADE-Grande, NEVOD-DECOR, Pierre Auger, SUGAR, Telescope Array, Yakutsk EAS Array} Collaboration, D.~Soldin {\em et~al.}, \href{http://dx.doi.org/10.22323/1.395.0349}{{\em PoS} {\bfseries ICRC2021} (2021) 349}.

\bibitem{Cazon:2020zhx}
{\bfseries EAS-MSU, IceCube, KASCADE Grande, NEVOD-DECOR, Pierre Auger, SUGAR, Telescope Array, Yakutsk EAS Array} Collaboration, L.~Cazon {\em et~al.}, \href{http://dx.doi.org/10.22323/1.358.0214}{{\em PoS} {\bfseries ICRC2019} (2020) 214}.

\bibitem{ArteagaVelazquez:2023fda}
J.~C. Arteaga~Velazquez, \href{http://dx.doi.org/10.22323/1.444.0466}{{\em PoS} {\bfseries ICRC2023} (2023) 466}.

\bibitem{Anchordoqui:2021ghd}
L.~A. Anchordoqui {\em et~al.}, \href{http://dx.doi.org/10.1016/j.physrep.2022.04.004}{{\em Phys. Rept.} {\bfseries 968} (2022) 1--50}.

\bibitem{Feng:2022inv}
J.~L. Feng {\em et~al.}, \href{http://dx.doi.org/10.1088/1361-6471/ac865e}{{\em J. Phys. G} {\bfseries 50} no.~3, (2023) 030501}.

\bibitem{Adhikary:2024nlv}
J.~Adhikary {\em et~al.}, \href{http://dx.doi.org/10.1140/epjc/s10052-025-14048-6}{{\em Eur. Phys. J. C} {\bfseries 85} no.~4, (2025) 430}.

\bibitem{FASER:2023zcr}
{\bfseries FASER} Collaboration, H.~Abreu {\em et~al.}, \href{http://dx.doi.org/10.1103/PhysRevLett.131.031801}{{\em Phys. Rev. Lett.} {\bfseries 131} no.~3, (2023) 031801}.

\bibitem{SNDLHC:2023pun}
{\bfseries SND@LHC} Collaboration, R.~Albanese {\em et~al.}, \href{http://dx.doi.org/10.1103/PhysRevLett.131.031802}{{\em Phys. Rev. Lett.} {\bfseries 131} no.~3, (2023) 031802}.

\bibitem{FASER:2024ref}
{\bfseries FASER} Collaboration, R.~Mammen~Abraham {\em et~al.}, \href{http://dx.doi.org/10.1103/PhysRevLett.134.211801}{{\em Phys. Rev. Lett.} {\bfseries 134} no.~21, (2025) 211801}.

\bibitem{FASER:2018ceo}
{\bfseries FASER} Collaboration, A.~Ariga {\em et~al.}, \href{http://dx.doi.org/10.48550/arXiv.1811.10243}{{\em CERN-LHCC-2018-030} (2018) }.

\bibitem{FASER:2020gpr}
{\bfseries FASER} Collaboration, H.~Abreu {\em et~al.}, \href{http://dx.doi.org/10.48550/arXiv.2001.03073}{{\em CERN-LHCC-2019-017} (2020) }.

\bibitem{dembinski_chromo_2023}
H.~Dembinski, A.~Fedynitch, and A.~Prosekin, \href{http://dx.doi.org/10.22323/1.444.0189}{{\em PoS} {\bfseries ICRC2023} (2023) 189}.

\bibitem{Riehn:2019jet}
F.~Riehn {\em et~al.}, \href{http://dx.doi.org/10.1103/PhysRevD.102.063002}{{\em Phys. Rev. D} {\bfseries 102} no.~6, (2020) 063002}.

\bibitem{Roesler:2000he}
S.~Roesler, R.~Engel, and J.~Ranft, \href{http://dx.doi.org/10.1007/978-3-642-18211-2_166}{{\em {International Conference on Advanced Monte Carlo for Radiation Physics, Particle Transport Simulation and Applications (MC 2000)}} (2000) 1033--1038}.

\bibitem{Pierog:2023ahq}
T.~Pierog and K.~Werner, \href{http://dx.doi.org/10.22323/1.444.0230}{{\em PoS} {\bfseries ICRC2023} (2023) 230}.

\bibitem{Ostapchenko:2013pia}
S.~Ostapchenko, \href{http://dx.doi.org/http://dx.doi.org/10.1051/epjconf/20125202001}{{\em EPJ Web Conf.} {\bfseries 52} (2013) 02001}.

\bibitem{Ostapchenko:2005nj}
S.~Ostapchenko, \href{http://dx.doi.org/10.1103/PhysRevD.74.014026}{{\em Phys. Rev. D} {\bfseries 74} no.~1, (2006) 014026}.

\bibitem{Ostapchenko:2024myl}
S.~Ostapchenko, \href{http://dx.doi.org/10.1103/PhysRevD.109.094019}{{\em Phys. Rev. D} {\bfseries 109} no.~9, (2024) 094019}.

\bibitem{Sjostrand:2014zea}
T.~Sj\"ostrand {\em et~al.}, \href{http://dx.doi.org/10.1016/j.cpc.2015.01.024}{{\em Comput. Phys. Commun.} {\bfseries 191} (2015) 159--177}.

\bibitem{Fieg:2023kld}
M.~Fieg {\em et~al.}, \href{http://dx.doi.org/10.1103/PhysRevD.109.016010}{{\em Phys. Rev. D} {\bfseries 109} no.~1, (2024) 016010}.

\bibitem{Riehn:2024prp}
F.~Riehn, A.~Fedynitch, and R.~Engel, \href{http://dx.doi.org/10.1016/j.astropartphys.2024.102964}{{\em Astropart. Phys.} {\bfseries 160} (2024) 102964}.

\bibitem{Anchordoqui:2022fpn}
L.~A. Anchordoqui {\em et~al.}, \href{http://dx.doi.org/10.1016/j.jheap.2022.03.004}{{\em JHEAp} {\bfseries 34} (2022) 19--32}.

\bibitem{Kling:2021gos}
F.~Kling and L.~J. Nevay, \href{http://dx.doi.org/10.1103/PhysRevD.104.113008}{{\em Phys. Rev. D} {\bfseries 104} no.~11, (2021) 113008}.

\bibitem{Nevay:2018zhp}
L.~J. Nevay {\em et~al.}, \href{http://dx.doi.org/10.1016/j.cpc.2020.107200}{{\em Comput. Phys. Commun.} {\bfseries 252} (2020) 107200}.

\bibitem{Andreopoulos:2015wxa}
C.~Andreopoulos {\em et~al.}, \href{http://dx.doi.org/10.48550/arXiv.1510.05494}{{\em FERMILAB-FN-1004-CD} (2015) }.

\bibitem{Buckley:2010ar}
A.~Buckley {\em et~al.}, \href{http://dx.doi.org/10.1016/j.cpc.2013.05.021}{{\em Comput. Phys. Commun.} {\bfseries 184} (2013) 2803--2819}.

\bibitem{Cruz-Martinez:2023sdv}
J.~M. Cruz-Martinez {\em et~al.}, \href{http://dx.doi.org/10.1140/epjc/s10052-024-12665-1}{{\em Eur. Phys. J. C} {\bfseries 84} no.~4, (2024) 369}.

\bibitem{Kling:2023tgr}
F.~Kling, T.~M\"akel\"a, and S.~Trojanowski, \href{http://dx.doi.org/10.1103/PhysRevD.108.095020}{{\em Phys. Rev. D} {\bfseries 108} no.~9, (2023) 095020}.

\end{thebibliography}\endgroup

\end{document}